\documentstyle[12pt,aasms4]{article}

\begin{document}

\title{Explosive Nucleosynthesis in Axisymmetrically Deformed
Type II Supernovae}

\author{Shigehiro Nagataki\altaffilmark{1}, Masa-aki
Hashimoto\altaffilmark{2}, Katsuhiko Sato\altaffilmark{1,3}, and
Shoichi Yamada\altaffilmark{1,4}}

\noindent
\altaffilmark{1}{Department of Physics, School of Science, the University
of Tokyo, 7-3-1 Hongo, Bunkyoku, Tokyo 113, Japan}\\
\altaffilmark{2}{Department of Physics, Faculty of Science,
Kyusyu University, Ropponmatsu, Fukuoka 810, Japan}\\
\altaffilmark{3}{Research Center for the Early Universe, School of
Science, the University of Tokyo, 7-3-1 Hongo, Bunkyoku, Tokyo 113, Japan} \\
\altaffilmark{4}{Max-Planck-Institute f\"ur Physik und Astrophysik
Karl-Schwarzschild Strasse 1, D-8046, Garching bei M\"unchen, Germany
}


\begin{abstract}
Explosive nucleosynthesis under the axisymmetric explosion in Type II supernova
has been performed by means of two dimensional hydrodynamical calculations.
We have compared the results with
the observations of SN 1987A. Our chief findings are as follows:
(1) $\rm ^{44}Ti$ is synthesized so much as to explain the tail of the 
bolometric light curve of SN 1987A. We think this is because the
alpha-rich freezeout takes place more actively under the axisymmetric
explosion.
(2) $\rm ^{57}Ni$ and $\rm ^{58}Ni$
tend to be overproduced compared with the observations. However,
this tendency relies strongly on the progenitor's model.

We have also compared the abundance of each element in the mass number range
$A= 16-73$ with the solar values. We have found three outstanding features.
(1) For the nuclei in the range $A=16-40$, their abundances are
insensitive to the initial form of the shock wave. 
This insensitivity is favored since the spherical
calculations thus far can explain the solar system abundances in this
mass range. 
(2) There is an enhancement around A=45 in the axisymmetric explosion
compared with the spherical explosion fairly well. In particular, $\rm
^{44}Ca$,
which is underproduced in the present spherical calculations, is enhanced
significantly.
(3) In addition, there is an enhancement around A=65. This tendency does
not rely on the form of the mass cut but of the initial shock wave.
This enhancement may be the problem of the overproduction in
this mass range, although this effect would be relatively small
since Type I supernovae are chiefly responsible for this mass number range.

\end{abstract}


\section{Introduction} \label{intro}

\indent

Supernovae play an important role in ejecting heavy elements produced in
massive stars (e.g., \cite{woosley95} and references therein).
It is important to determine
the composition of ejected gas as a function of stellar mass since it
is basic data for the chemical evolution of galaxies. 
In this paper, we discuss Type II
supernovae, which are regarded as the death of a massive star whose
mass exceeds 8 times the solar mass ($M_{\odot}$) (e.g., \cite{hashimoto95}).

The mechanism of Type II supernovae has been understood as follows
(e.g., \cite{bethe90}): when the
mass of the iron core of the progenitor exceeds the Chandrasekhar
mass, the star begins
to collapse. The collapse continues until the central density of
the collapsing core reaches about 1.5-2 times the nuclear matter
density ($\rho = 2.7 \times 10^{14} \rm g\ cm^{-3}$),
beyond which matter becomes too stiff to be compressed further. A shock
wave then forms, propagates outward. At first, the shock wave is not so strong
and stall in the Fe core. However, by the neutrino heating, 
the shock wave is revived, begins to propagate outward again, 
and finally produces the supernova explosion.
This phenomenon is called as delayed explosion, which is the most promising
theory for the mechanism of type II supernova explosions.

When the shock wave passes Si-- rich and O-- rich layers,
the temperature becomes high
enough to cause many nuclear reactions.  This phenomenon is called the 
explosive nucleosynthesis in Type II supernovae.
Many calculations have so far been
performed on explosive nucleosynthesis in supernovae
(e.g., \cite{woosley86}; \cite{hashimoto89}; \cite{thielemann90};
\cite{hashimoto95}, \cite{woosley95}).

It is SN 1987A in the Large Magellanic Cloud that has provided the
most precise data to test the validity of such calculations.
For example, the bolometric luminosity began to increase in a few
weeks after the explosion (\cite{catchpole87}; \cite{hamuy87}), 
which is attributed to the decay of the radioactive nucleus $^{56}
\rm Ni$. $^{56} \rm Ni$ is
synthesized during the explosion, and the mass is estimated to be $0.07-0.076
M _{\odot}$ on the basis of the luminosity study (\cite{shigeyama88};
\cite{woosley88}).
$^{57} \rm Ni$ and $^{44} \rm Ti$ are also thought to be important
nuclei to explain the bolometric light curve. 
Since the half lives of these nuclei are longer than that of 
$^{56}\rm Co$, their decays are thought to be responsible for the tail
of the light curve. In fact, the observed bolometric light curve's decline rate
is slowed down since $\sim$ 900 days after explosion (\cite{suntzeff91}).
The ratio of $^{57} \rm Ni$ to $^{56} \rm Ni$ is estimated from the X-ray
light curve to be $1.5 \pm 0.5$ times the solar $\rm
^{57}Fe$/$\rm ^{56}Fe$ ratio (\cite{kurfess92}).
The ratio $\rm
^{44}Ti$/$\rm ^{56}Ni$ is also estimated and must be larger than 1.8
times the solar $\rm
^{44}Ca$/$\rm ^{56}Fe$ ratio if the contribution from the pulsar is
negligible (\cite{kumagai93}). 
There is other important nucleus whose amount is estimated by the
observation of
SN 1987A. That nucleus is $\rm ^{58}Ni$, which is produced at the
innermost region of the ejecta and gives very important information
about the mass cut.
From the spectroscopic
observation of SN 1987A, the ratio $<\rm ^{58}Ni/^{56}Ni>\equiv[\it
X(\rm ^{58}Ni)/ \it X(\rm ^{56}Ni)]/[\it X(\rm ^{58}Ni)/\it X(\rm
^{56}Fe)]_{\odot}$ should be 0.7-1.0 (\cite{rank88}).

Numerical calculations can reproduce the amount of $\rm ^{56}Ni$
, the ratio of $^{57} \rm Ni$ to $^{56} \rm Ni$, and the ratio of
$^{58} \rm Ni$ to $^{56} \rm Ni$ (\cite{hashimoto89}). However, 
the ratio of $\rm ^{44}Ti$ to $^{56} \rm Ni$ has never been reproduced. 
It is reported that $\rm ^{44}Ti$ cannot be produced enough to explain the
tail of the light curve in wide parameter range (\cite{woosley91}).
Pulsar is another candidate that can explain the light curve. If the
energy supply by the pulsar dominates, the light curve will be flat.
However, such flatness has not been observed yet.     
The effect of long recombination and cooling time scales of the
remnant is also considered for the explanation (\cite{fransson93}).
However, they admit that their results depend on the model of the 
progenitor and more careful calculation must be needed for the 
quantitative estimates.
In the present circumstances,
the explanation of the bolometric light curve is still open to 
argument.

There is another touchstone of the numerical simulations. It is the
solar system abundances. The appropriate combination of the
contribution from  Type I and Type II supernova can reproduce the
solar system abundance ratios within a factor of 2 for typical
species (\cite{hashimoto95}).
However, there are some problems. For example, $\rm ^{35}Cl$, 
$\rm ^{39}K$, and $\rm ^{44}Ca$ are synthesised only about one tenth of
the solar value. On the other hand, $\rm ^{58} Ni$ is produced about
three times.

Are there any effects which can solve the problems mentioned above, that
is, $\rm ^{44}Ti$ problem in SN 1987A and reproduction of the solar system
abundance ? We suggest the effect of asymmetric (in particular,
axisymmetric) explosion will change the present circumstances.  
All calculations about nucleosynthesis have been done on the
assumption that the explosion is spherically symmetric. However, there are
some reasons we should take account of the asymmetry in
supernova explosion. Among them is a well-known fact that most massive
stars are rapid rotators (\cite{tassoul78}). Since stars are rotating
in reality, the effect of rotation should be investigated in numerical
simulations of a collapse-driven supernova. Thus far, several
simulations have been done by a few groups in order to study rotating
core collapse (\cite{muller80}; \cite{tohline80}; \cite{muller81};
\cite{bodenheimer83}; \cite{symbalisty84};
\cite{mochmeyer89}; \cite{finn90}; and
\cite{yamada94}).
As a result, some numerical simulations of a collapse-driven supernova
suggest the possibility of axisymmetric explosion if the effect of a stellar
magnetic field and/or stellar rotation is taken into
consideration.
There is also possibility that the axisymmetrically modified neutrino
radiation from a rotating proto-neutron star causes asymmetric explosion
(\cite{shimizu94}). We note these effects mentioned above tend to cause
axisymmetric explosion. 
Furthermore, many observations of SN 1987A suggest
the asymmetry of the explosion. The clearest is the speckle images of
the expanding envelope with high angular resolution
(\cite{papaliolis89}), where an oblate shape with
an axis ratio of $\sim 1.2 - 1.5 $ was shown.  Similar results were also
obtained from the measurement of the linear polarization of the scattered
light from the envelope (\cite{cropper88}). If the envelope is
spherically symmetric, there is no net linear polarization
induced by scattering. Assuming again that the shape of the scattering
surface is an oblate or prolate spheroid, one finds that the observed
linear polarization corresponds to an axis ratio of $\sim 1.2$.

Because of the reason mentioned above, it is important to investigate
the effect of axisymmetric explosion on explosive nucleosynthesis.
In the present paper, we calculate the explosive nucleosynthesis for a
20$M_{\odot}$ star under the axisymmetric explosion and
investigate if the difficulties mentioned above are improved by its effect.

We show our method of calculation for the explosive nucleosynthesis
in section \ref{calculation}. Results are presented in section
\ref{results}. Summary and discussion are given in section \ref{summary}.

\section{ Model and Calculations } \label{calculation}

\subsection{ Hydrodynamics} \label{hydro}
\indent

We performed 2-dimensional hydrodynamical calculations.
The calculated region corresponds to a quarter part of the meridian
plane under the assumption of axisymmetry and equatorial symmetry.
The number of meshes is $ 300 \times 10$
(300 in the radial direction, and 10 in the angular direction). 
The inner and outer most radius are set to be $10^{8} \rm cm$ and $2
\times 10^{10} \rm cm$, respectively.
We use the Roe method for the calculation (\cite{roe81}; \cite{yamada94}).
The basic equations are as follows:
\begin{eqnarray*}
\partial_{t} \rho =&& -\frac{1}{r^2}\partial_{r}(\rho u_{r} r^2)
 -\frac{1}{r\sin \theta}\partial_{\theta}(\rho u_{\theta}\sin \theta), \\
\partial_{t} (\rho u_{r}) =&& -\frac{1}{r^2}\partial_{r}(\rho u_{r}^2 r^2)
-\frac{1}{r\sin \theta}\partial_{\theta}(\rho u_r u_{\theta}\sin \theta) \\
&& -\partial_rP + \frac{\rho u_{\theta}^2}{r}, \\
\partial_{t} (\rho u_{\theta}) =&& -\frac{1}{r^2}\partial_r(\rho u_{\theta} u_rr^2) -\frac{1}{r\sin \theta}\partial_{\theta}(\rho u_{\theta}^2\sin \theta)\\
&& -\frac{1}{r}\partial_{\theta}P-\frac{\rho u_{\theta} u_r}{r} ,\\
\partial_tE =&& -\frac{1}{r^2}\partial_r\left[(E+P)u_r r^2\right] \\ 
&& -\frac{1}{\sin \theta}\partial_{\theta}\left[(E+P) u_{\theta}\sin \theta \right]
\\
\end{eqnarray*}
where $\rho, P,$ and $E$ are the mass density,
pressure, total energy
density per unit volume and $u_{r}$ and $u_{\theta}$ are velocities of 
a fluid in $r$ and $\theta$ direction, respectively.
The first equation is the continuity equation, the second and third are 
the Euler
equations and the forth is the equation of the energy conservation.
We use the equation of state:
\begin{eqnarray*}
P = \frac{1}{3} a T^{4} + \frac{\rho k_B T}{ A_{\mu} m_u }
\end{eqnarray*}
where $a$, $ k_B$, $ A_{\mu}$ and $ m_u$
are the radiation constant, Boltzmann
constant, the mean atomic weight, and the atomic mass unit,
respectively.

In this paper, we assume the system is adiabatic after the passage of
the shock wave,
because the entropy produced during the explosive nucleosynthesis is
much smaller than that generated by the shock wave. As a result, the
entropy per nucleon is conserved. 

\subsection{ Post-processing} \label{particle}

\indent

In order to calculate the change of the chemical composition of the
star, we use a test particle approximation. $(10(r)
\times 10(\theta))$ and $(40(r) \times 10(\theta))$ particles are 
scattered in the Si-- rich and O--
rich layers, respectively, with the increasing interval in the radial
directions and same interval in the angular ones in each layer. This
is because explosive nucleosynthesis can occur mainly in the inner region
where the temperature reach high enough to cause nuclear reactions.
On the other hand, little change occurs in the outer region and we do
not need to scatter many particles in the outer region.

We preserve the time evolution of 
density and temperature along each trajectory of test particles. 
It is assumed that test particles are at rest at first and move with
the local velocity at their positions after the passage of a shock
wave. Thus we can calculate each particle's path by integrating 
$\displaystyle{\frac{\partial \vec{x}}{\partial t}} 
= \vec{v}(t, \vec{x}) $, where
the local velocity $\vec{v}(t, \vec{x})$ is given from the
hydrodynamical calculations mentioned above. The density and
the temperature of a test particle at each time are determined by
interpolation in the Eulerian mesh the particle is in at the moment. 
Nucleosynthesis calculations are done separately for each trajectory 
of test particles using a nuclear reaction network explained below (after the
hydrodynamical simulations). 
In calculating the total yields of elements we assume that each
test particle has its own mass which is determined from the initial 
distribution of the test particles so that their sum becomes 
the mass of the Si-- rich and O-- rich layers, and also assume 
that the nucleosynthesis occurs uniformly in each mass element. 
In this way, the total chemical composition can be calculated by
summing the final chemical composition of each mass element weighted 
by its mass. 

It is noted that the above assumption is valid when the typical size 
of a test particle calculated from its mass and density is smaller
than the temperature scale height at its position. It is also
necessary that the shear of the flow is not very large in the mass
element. In this calculation we assumed that the initial velocity 
of the matter is radial behind the shock wave (see $\S$\ref{init}) 
and the time scale of the explosive nucleosynthesis is small ($\sim$ 1 sec), we
think this assumption is not so bad.
Regardless, this is only  
the first step of the estimation of chemical abundances in an 
axisymmetric supernova explosion, and improvement in resolution of 
mesh and test particles is now underway.

\subsection{ Nuclear reaction network}

\indent

\begin{figure}
\epsscale{1.0}
\plotone{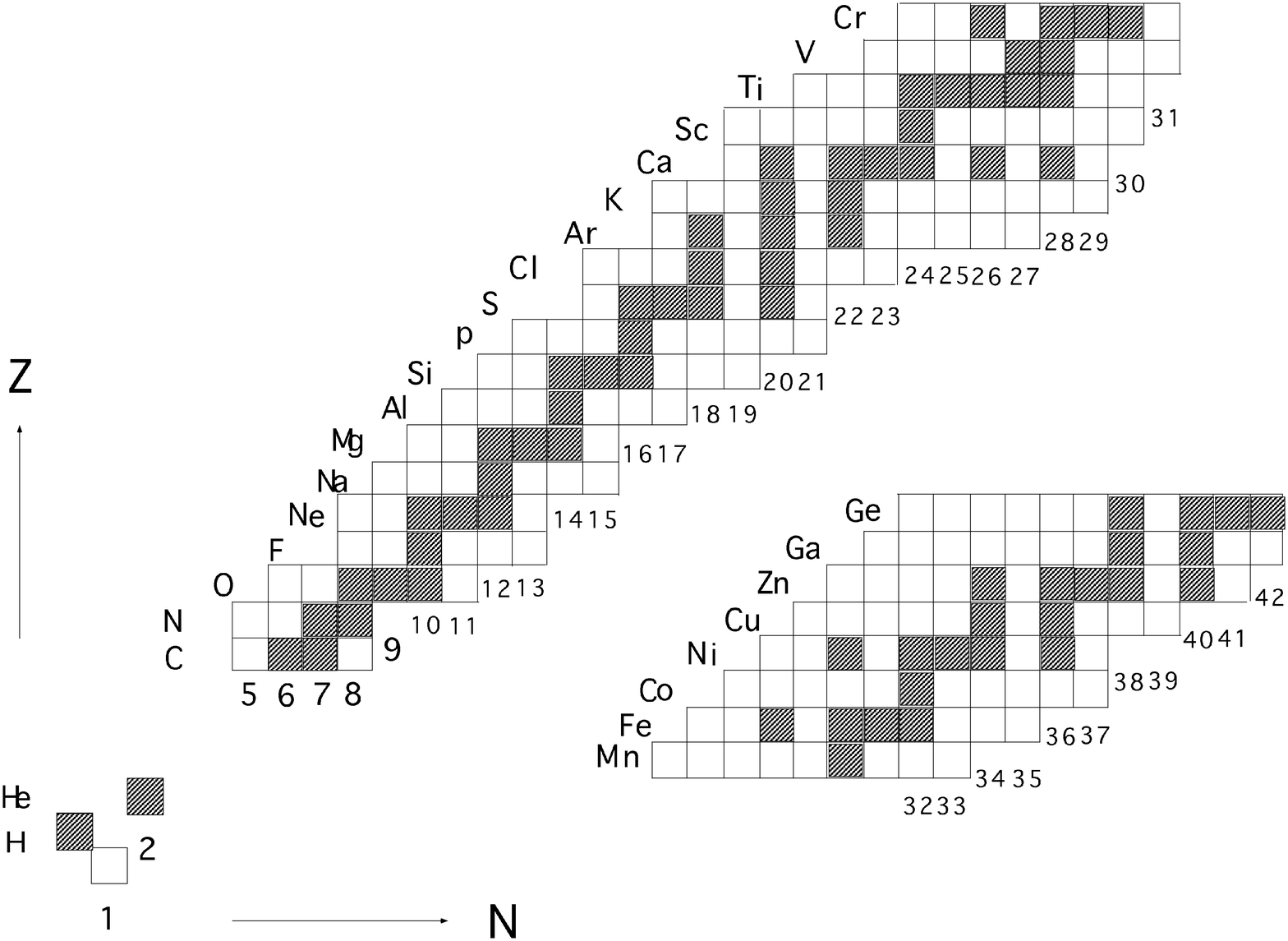}
\figcaption{Table of nuclei included in our nuclear reaction network.
242 species are included. The gray-colored nuclei denote stable nuclei.
 \label{fig1}}
\end{figure}

We have calculated the explosive nucleosynthesis using the time evolution 
of $(\rho,T)$ discussed in $\S$\ref{particle}. Since the system is not in
chemical equilibrium, we must calculate the change of the chemical
composition with the use of the nuclear reaction network. It contains 
242 species (see Figure~\ref{fig1}, \cite{hashimoto89}). The basic
equations for the abundance changes are
\begin{eqnarray*}
\partial_t y_{i}=\alpha_{ijkl}y_{j}y_{k}y_{l}+\beta_{ijk}y_{j}y_{k}
+\gamma_{ij}y_{j}, \\
y_{i}=n_{i}/(\rho N_{A}),
\end{eqnarray*}
where $n_{i}$ is the number density of i-th nucleus $i$ and $N_{A}$ is
Avogadro's number. Reaction rates are defined by $\alpha_{ijkl}$,
$\beta_{ijk}$, and $\gamma_{ij}$ which have dimensions of $\rm sec^{-1}$.
The first term of the right hand side represents 3-body reactions such
as the triple-alpha process, the second is for 2-body, and the third
is for 1-body such as the photo-disintegration or $\beta$ decay.
We integrate this system of coupled differential equations by an
implicit method (\cite{hashimoto83}). To construct a large network
relevant for the calculation of the explosive nucleosynthesis, data
of reaction rates were used from various sources (e.g., \cite{hashimoto85}). 
\placefigure{fig1}

\subsection{ Initial conditions} \label{init}

\indent

The progenitor of SN 1987A, Sk-69$^{\circ}$202, is thought to
have had the mass $\sim 20 M_{\odot}$ in the main-sequence stage
(\cite{shigeyama88}; \cite{woosley88}) and had $\sim$ (6$\pm$1)$M_{\odot}$
helium core (\cite{woosley88a}).
In the present paper, the presupernova model which is obtained 
from the evolution of a helium core of 6 $M_{\odot}$ (\cite{nomoto88})
is used for the initial density and composition.
Table \ref{radius} shows the radii of the Fe/Si, Si/O and O/He
interfaces in this model.

We will explain the initial shock wave.
Since there is still uncertainty as to the mechanism of type II supernova,
precise explosive nucleosynthesis calculations have not been performed from
the beginning of core collapse. Instead, 
explosion energy is deposited artificially at the innermost boundary (e.g., 
\cite{hashimoto95}). There is another method used by Woosley
(``the method of the piston'', see \cite{woosley95}). However, both
methods are only approximation and the initial condition has been problem of 
the explosive nucleosynthesis.
In this paper, the method of energy deposition is taken and
the explosion energy of $1.0 \times 10^{51} \rm
erg$ is injected to the region from $1.0 \times 10^{8} \rm cm$ to $1.5
\times 10^{8} \rm cm $ (that is, at the Fe/Si interface).

As for the axisymmetric explosion, 
the initial velocity of matter
behind the shock wave is assumed to be
radial and proportional to $r \times \frac{1+\alpha \cos(2 \theta)}
{1+ \alpha }$, where r, $\theta$, and $\alpha$ are radius, the zenith
angle, and the free parameter which determine the degree of the
axisymmetric explosion, respectively. 
Since the ratio of the velocity in the polar region to that in the
equatorial region is 1 : $\frac{1-\alpha}{1+\alpha}$, 
more extreme jet-like shock waves is obtained as the $\alpha$ gets larger.
In the present study, we take $\alpha=0$ for the spherical explosion 
and $\alpha=\frac{1}{3},\; \frac{3}{5}, \; \rm and \; \frac{7}{9}$
(these values mean that the ratios of the velocity are 2:1, 4:1, and 8:1,
respectively) for the axisymmetric ones (see Table \ref{model}).
We assumed that the distribution of thermal energy is same as the velocity
distribution and that total thermal energy is equal to total kinetic energy.

We note that the form of the initial shock wave can not be known
directly from both observation and theory. As a result of it, the value of
$\alpha$ can not be known $\it a \; priori$.
However, it is reported that
there is a possibility for a shock wave to be
jet-like if the proper angular momentum of the progenitor is assumed
(\cite{yamada94}).
Because of this reason, we think our formulation of the
initial shock wave is not so unreasonable. At least, there is a
possibility for the shock wave to be axisymmetric as we have assumed.
Moreover, the value of $\alpha$ to be desired in SN 1987A
would be in this range as shown in section \ref{results}.

\placetable{radius}
\placetable{model}

\section{ Results} \label{results}

\indent

\subsection{Reproduction of observational data of SN 1987A} \label{1987A}

\subsubsection{Mass cut}

In type II supernova, there is boundary that separates ejecta and
central compact object. This boundary is called as the mass cut.
Strictly speaking, the position of the mass cut should be determined
by hydrodynamical calculation including gravity, that is, the matter
which has positive total energy (sum of the kinetic, thermal, and
gravitational energy) can escape and that which has negative energy
falls back to central compact object. 
However, it is very difficult to determine the position of the mass cut
hydrodynamically since it is sensitive not only to the explosion
mechanism, but also to the presupernova structure, stellar mass, and
metallicity. In fact, it is reported that total amount of $\rm
^{56}Ni$ can not be reproduced by the piston method, which determines the 
mass cut hydrodynamically (\cite{woosley95}). 

\begin{figure}
\epsscale{1.0}
\plottwo{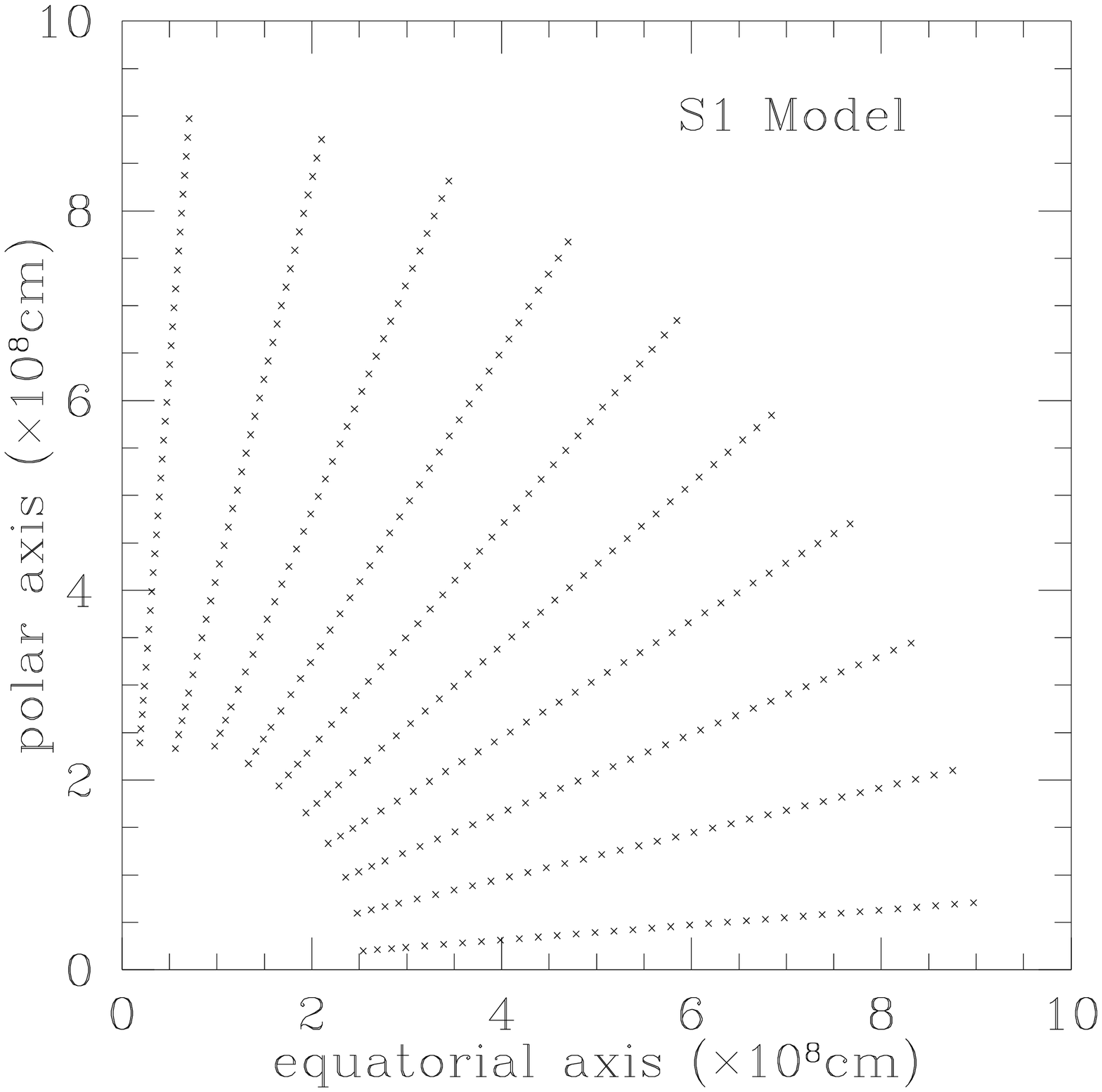}{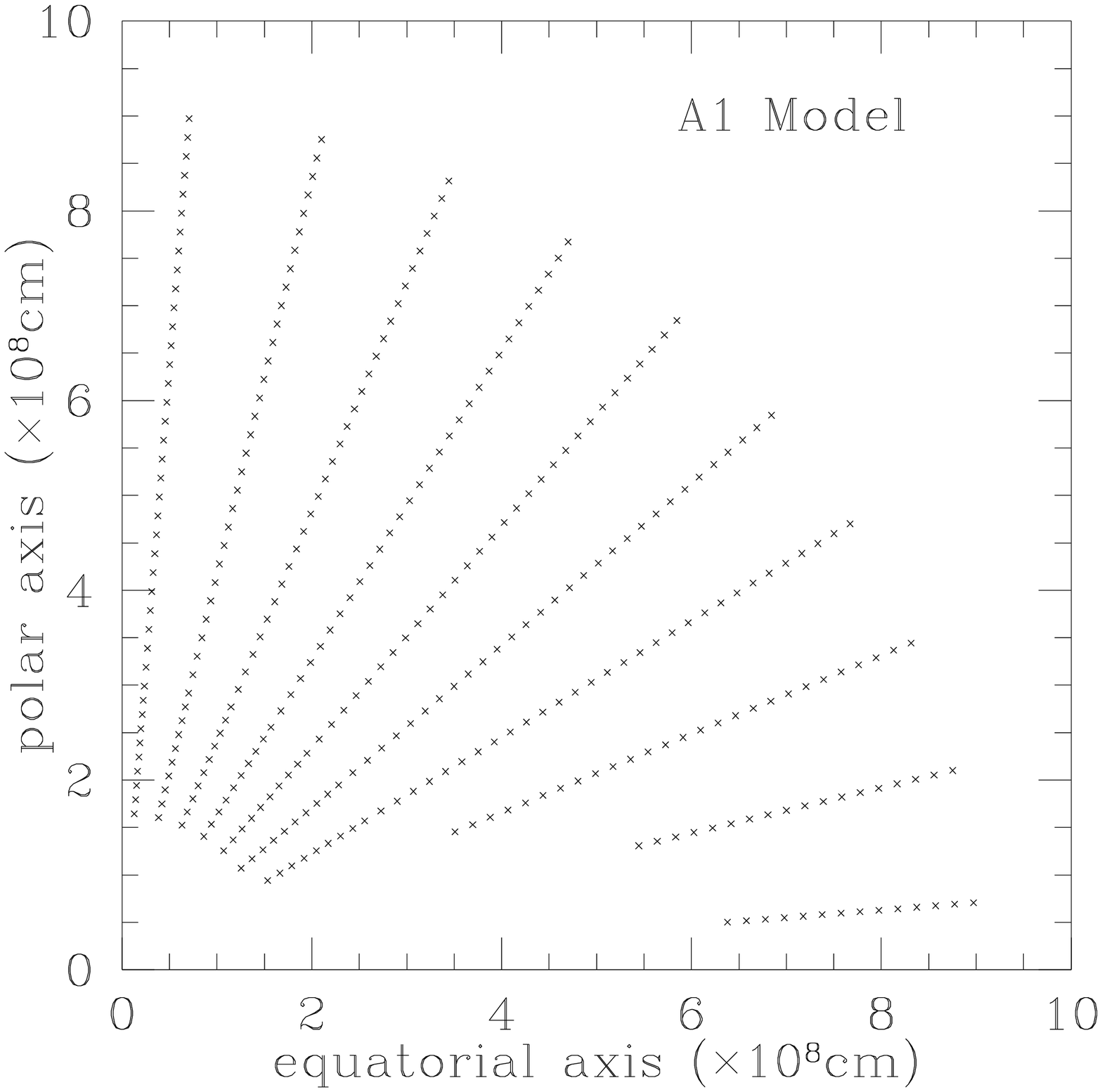}
\figcaption{Form of the mass cut (A7 model) for S1 case and A1 case. 
Dots are plotted for the initial positions of the test particles which 
will be ejected.
\label{fig2}}
\end{figure}

\begin{figure}
\epsscale{1.0}
\plottwo{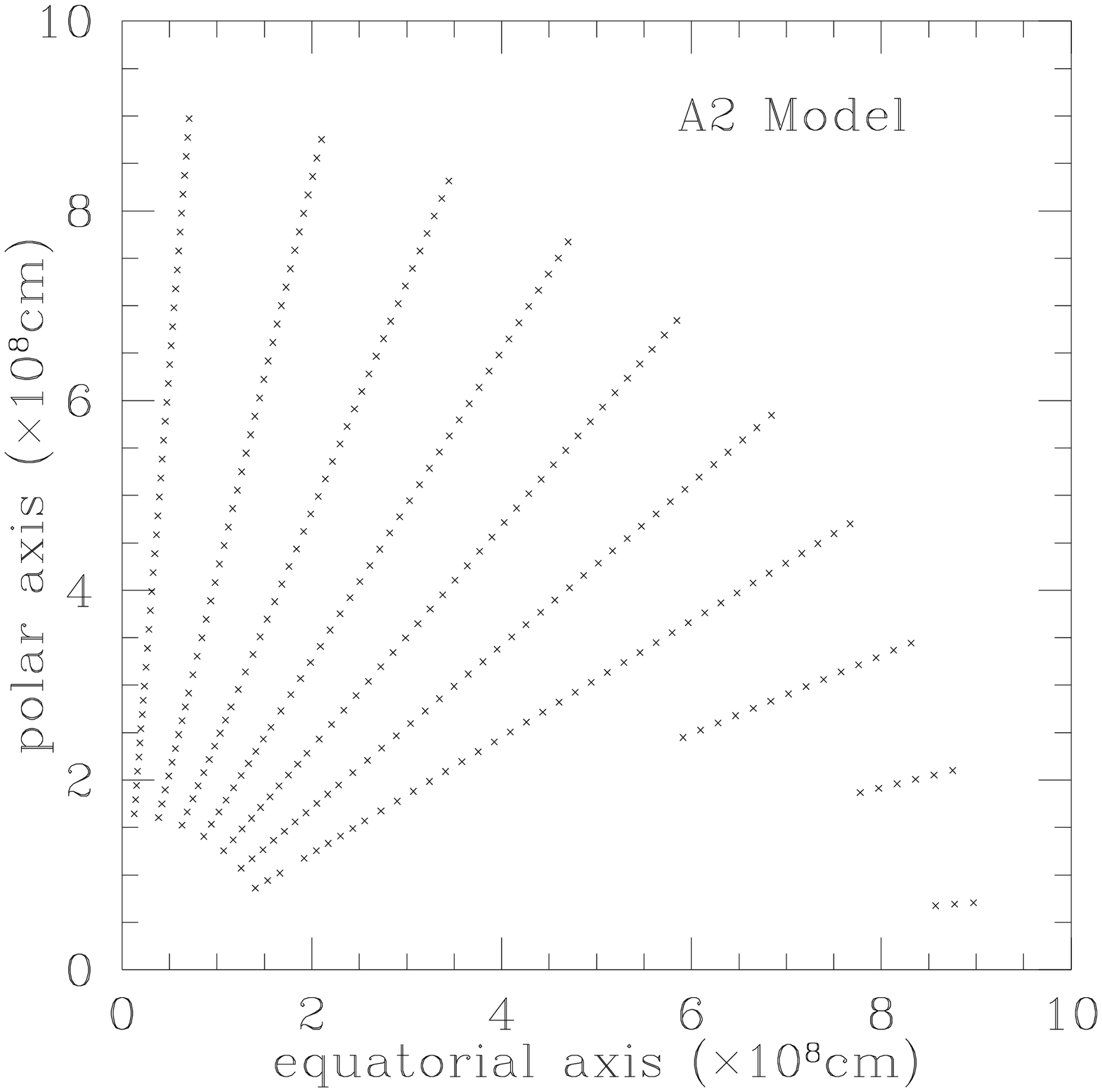}{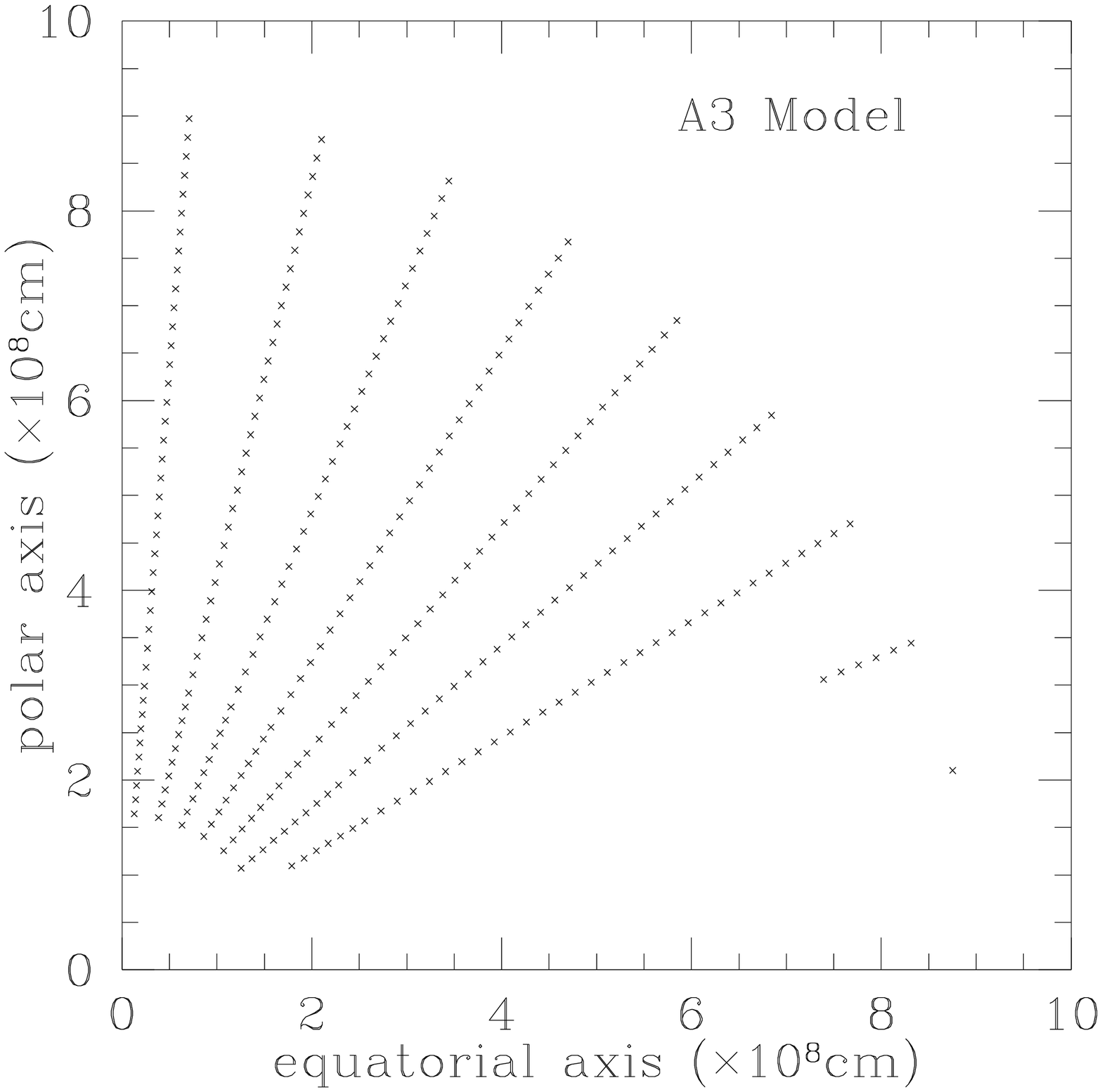}
\figcaption{Same as Fig.2 but for A2 and A3 case.
\label{fig3}}
\end{figure}

There is another way to determine the position of the mass cut.
Among many observational data of SN 1987A, the total amount of $\rm ^{56}Ni$ in
the ejecta is one of the most reliable one.
For that reason, the position of the mass cut can be
determined so as to contain $\sim 0.07 M_{\odot}$ $\rm ^{56}Ni$ in the 
ejecta (\cite{hashimoto95}). We took the same way in this paper.
However, this
method is simple only for spherical calculations. We must extend this
method for multi-dimensional calculations as follows:
we assume that the larger total energy (internal energy plus kinetic
energy) a test particle has, the more favorably it is
ejected (\cite{shimizu93}).
We first calculate the total energy of each test particle at the final
stage of our calculations ($\sim$ 10 sec) and then add up
the mass of $\rm ^{56}Ni$ in a descending order of the total energy
until the summed mass reaches $0.07  M _{\odot}$. The rest of
$\rm^{56}Ni$ is assumed to fall back to the central compact object even
if it has a positive energy at that time.
We show the position of the mass cut for
each model in Figures~\ref{fig2} and ~\ref{fig3}. We note the
dots, which show test particles which will be ejected, are plotted
for their initial positions. In this way, the position of the mass cut 
is easily determined.  
The tendency that more matter around the polar axis is ejected is
consistent with the initial form of the shock wave. We refer to this mass
cut as A7.
To see the dependence of our analysis on the position of the mass cut, 
we take another mass cut for comparison. This mass cut is set to be
spherical and determined so as to contain $0.07 M_{\odot}$ $\rm
^{56}Ni$ in the ejecta. We refer to this mass cut as S7.

\placefigure{fig2}
\placefigure{fig3}

\subsubsection{Comparison with SN 1987A} \label{with1987a}

\begin{figure}
\epsscale{1.0}
\plottwo{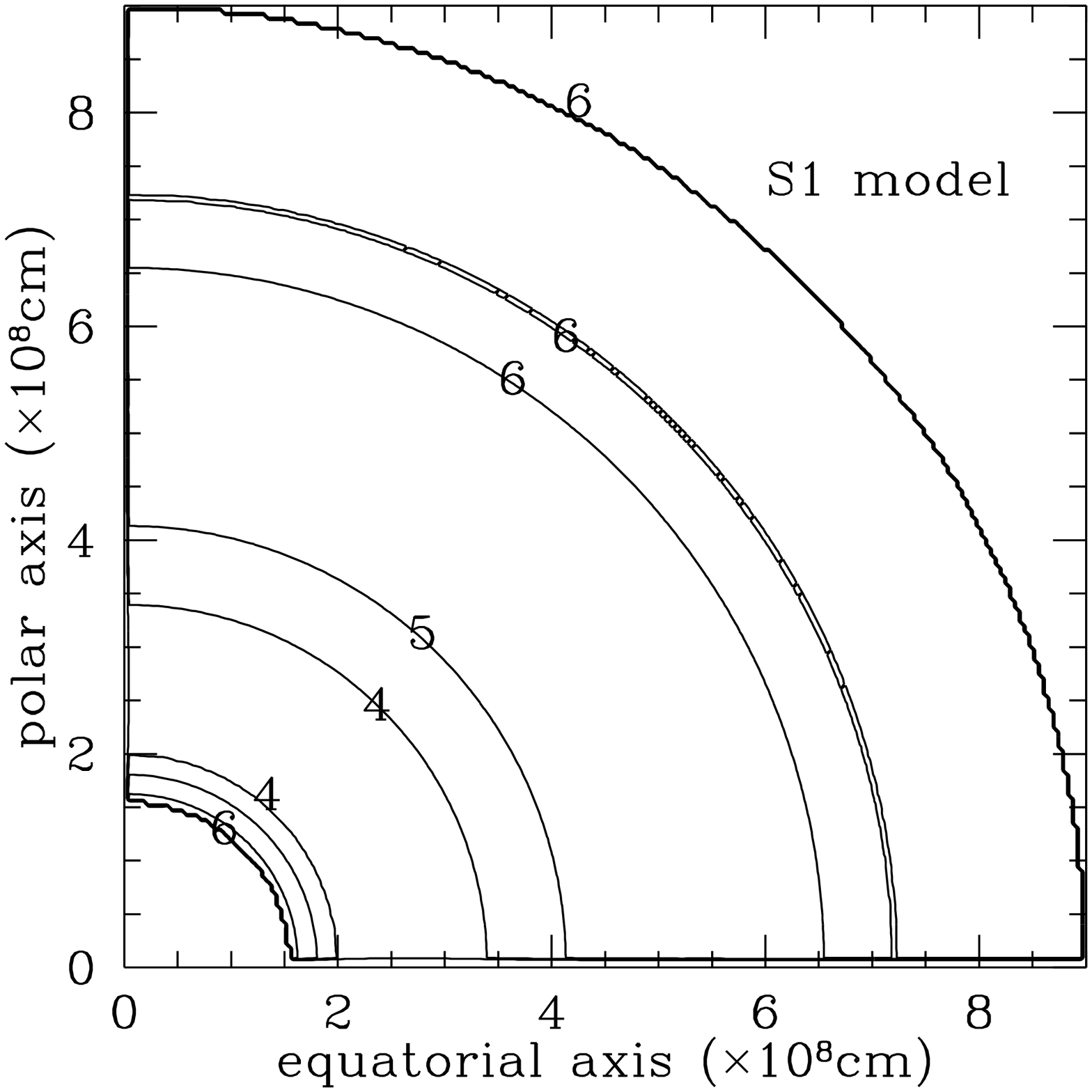}{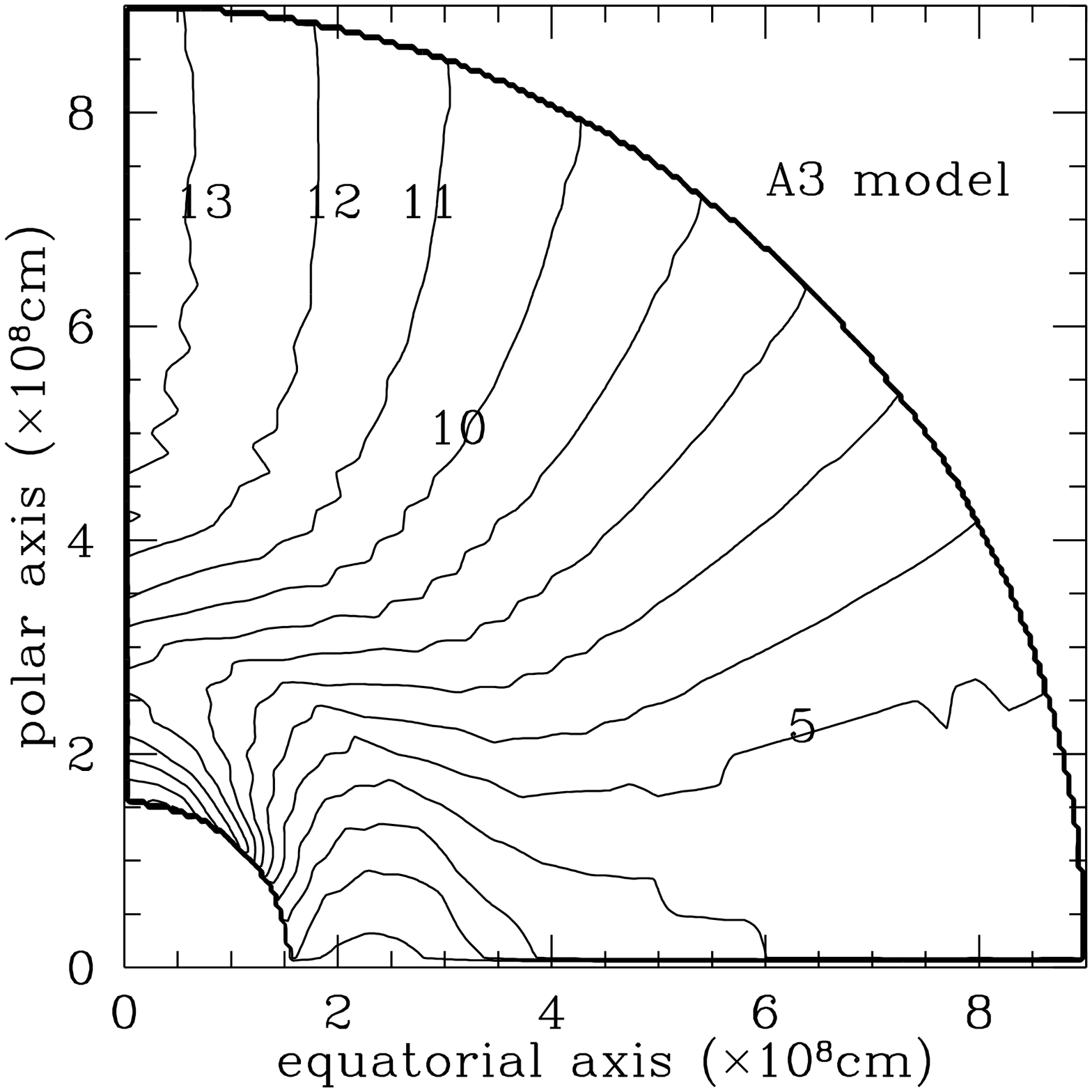}
\figcaption{
Distributions of the entropy for the two cases (left : S1 model, right 
: A3 model). Entropy is normalized by $k_B$. Contours are drawn for the
initial position of test particles.
\label{fig4}}
\end{figure}

\begin{figure}
\epsscale{1.0}
\plottwo{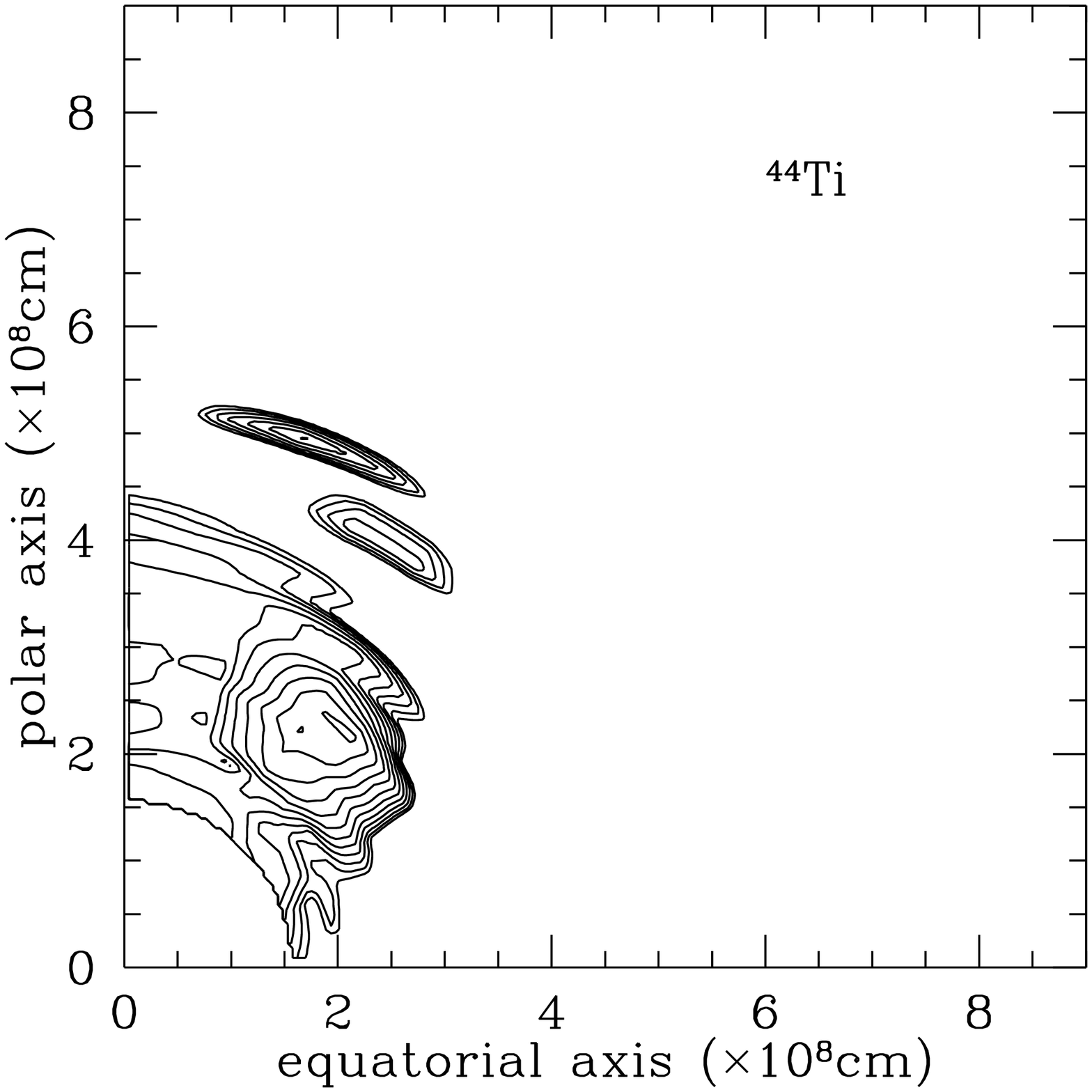}{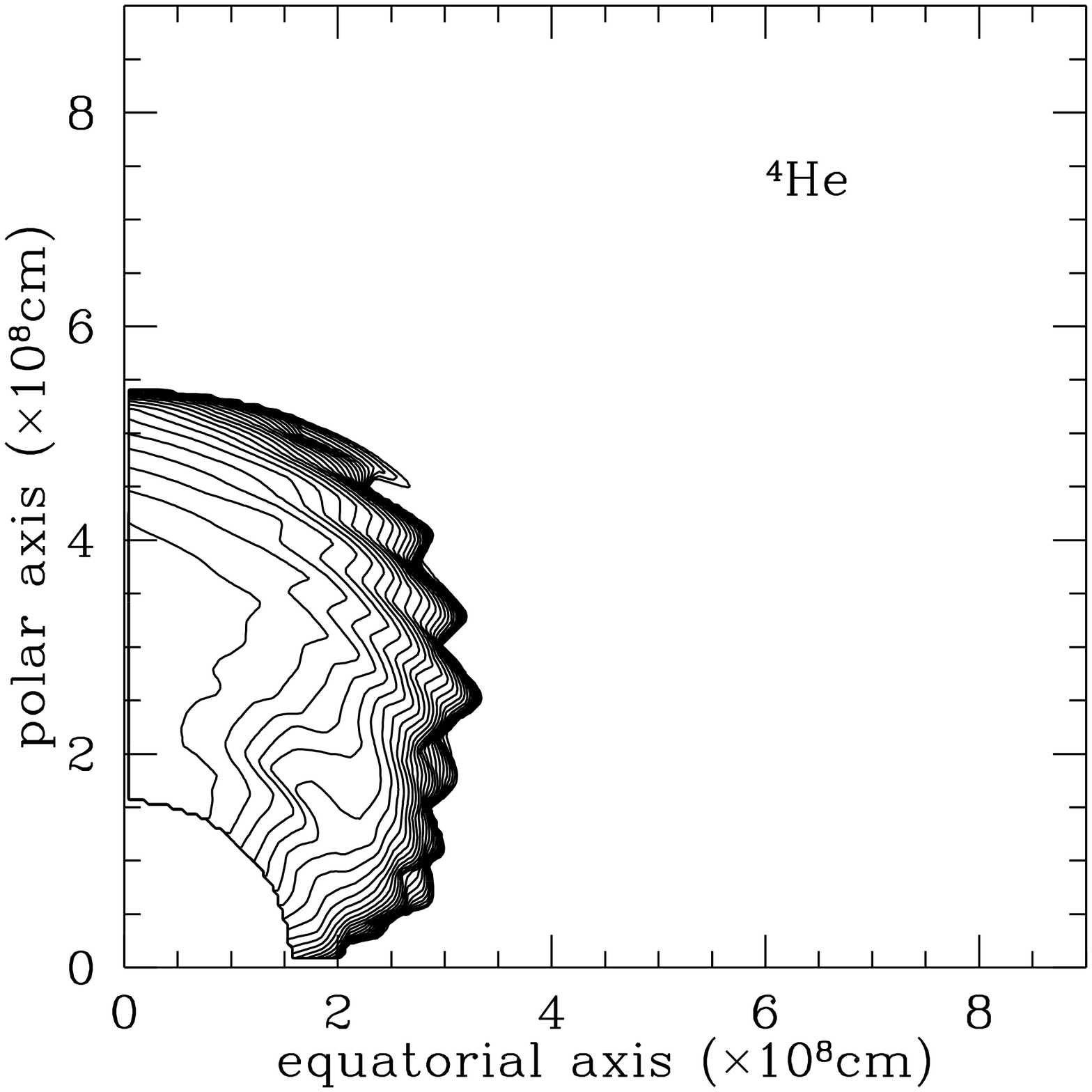}
\figcaption{Left: Contaur of the mass fraction of $^{44}\rm Ti$ in the
A3 case. The maximum value of the mass fraction of$^{44}\rm Ti$ is
1.3$\times 10^{-2}$. Right : Same as left but for $\rm ^{4} He$. The
maximum value is 4.0$\times 10^{-1}$.   
Contours are drawn for the initial position of test particles.
 \label{fig5}}
\end{figure}

We show the ratios $<\rm ^{44}Ti/^{56}Ni>$, $<\rm ^{57}Ni/^{56}Ni>$,
and $<\rm ^{58}Ni/^{56}Ni>$.
These quantums are defined as below:

\begin{eqnarray*}
<\rm ^{44}Ti/^{56}Ni>\equiv[\it X(\rm ^{44}Ti)/ \it X(\rm
^{56}Ni)]/[\it X(\rm ^{44}Ca)/\it X(\rm ^{56}Fe)]_{\odot}  \\
<\rm ^{57}Ni/^{56}Ni>\equiv[\it X(\rm ^{57}Ni)/ \it X(\rm
^{56}Ni)]/[\it X(\rm ^{57}Fe)/\it X(\rm ^{56}Fe)]_{\odot}  \\
<\rm ^{58}Ni/^{56}Ni>\equiv[\it X(\rm ^{58}Ni)/ \it X(\rm
^{56}Ni)]/[\it X(\rm ^{58}Ni)/\it X(\rm ^{56}Fe)]_{\odot}  \\
\end{eqnarray*}
where X denotes mass fraction. At first, there is a fact to which we must 
pay attention. 
It is reported that this 6 $M_{\odot}$ model is neutron-rich and the
value of $Y_e$ for $M > 1.607 M_{\odot}$ (=0.494) is artificially
changed to that of $M > 1.637 M_{\odot}$ (=0.499) to suppress the
overproduction of neutron-rich nuclei (\cite{hashimoto95}). 
This means the range of convective mixing in the presupernova model is 
artificially changed. Anyway, at first, we also modify
the 6 $M_{\odot}$ model in the same way. The result is summarised in
Table \ref{light} (Case A in the Table).
We can see clearly that $\rm ^{44}Ti$ is more produced as the
degree of the axisymmetric explosion gets larger. It is also noted that
$\rm ^{44}Ti$ is produced so much in the axisymmetric explosion 
as to explain the tail of the bolometric light curve of SN 1987A.
Since $\rm ^{44}Ti$ is synthesised through the alpha-rich freezeout, high
entropy is needed for the synthesis of this nucleus. Since the matter
becomes radiation dominated after the passage of the shock wave, the
entropy per baryon can be written approximately as below:
\begin{eqnarray*}
S_{\gamma} = \frac{16 \sigma}{3 k_B c} m_u  \frac{T^3}{\rho N_A}
\end{eqnarray*}
where $\sigma$, $k_B$, c, $m_u$, T, $\rho$, and $N_A$ are
Stefan-Boltzmann constant, Boltzmann constant, speed of light, atomic
mass unit, temperature, density, and Avogadro constant, respectively.
The entropy is normalized in unit of $k_B$.
We show distributions of
the entropy per nucleon for S1 and A3 models in Figure~\ref{fig4}.
We note the contours are drawn for the initial positions of the test particles.
It is clear from Figure~\ref{fig4} that higher entropy is achieved in the
polar region for the axisymmetric explosion. 
We will explain this tendency.
The relation between energy density and temperature behind the shock wave
can be written approximately as below:
\begin{eqnarray*}
E [erg/cm^3]= a T^4
\end{eqnarray*}
where a is radiation constant. As more energy is deposited initially in the
polar region, higher temperature is achieved. As a result of it,
higher entropy is achieved in the polar region than the equatrial
region.
We show in Figure ~\ref{fig5}
the contour of $\rm ^{44}Ti$ and $\rm ^{4} He$ in A3 model. 
$\rm ^{44}Ti$ is produced much
in the polar region together with $\rm ^{4} He$, as expected.

\placefigure{fig4}
\placefigure{fig5}

On the other side, $\rm ^{57}Ni$ and $\rm ^{58}Ni$ are overproduced
and inconsistent with the observations in the axisymmetric explosion.
This is because the ejecta contains the neutron-rich matter in the
polar region for the
axisymmetric explosion cases (see Fig ~\ref{fig2} and ~\ref{fig3}).
We note that even if the mass cut is set to be
spherical, the ejecta contains more neutron-rich matter in the
axisymmetric case.
This is because the mass cut tends to be smaller in axisymmetric
explosion since $\rm ^{56}Ni$ is less produced (see the mass cut of S7 
in Table \ref{light}). 
To see the effect of neutron-rich matter, we perform the 
same calculations for the modified 6 $M_{\odot}$ model, in which
the value of $Y_e$ for $M > 1.5 M_{\odot}$ is artificially
changed to that of $M > 1.637 M_{\odot}$. The results are summarised
in Table \ref{light} (Case B in the Table). We can see the
amount of $\rm ^{44}Ti$ is still
much enough to explain the light curve, with the amount of $\rm
^{57}Ni$ and $\rm ^{58}Ni$ consistent with the observations. 
This means that the
uncertainty of the presupernova model have a great influence on the
chemical composition of the ejecta.

We will give an additional comment on the mass cut and the mass of the 
central compact object. There is a tendency that as the degree of the
axisymmetric explosion gets larger, the mass of the central compact
object also becomes larger if the mass cut is set to be A7 (see Table
\ref{light}). In particular, there is a possibility that the mass of
the central object is large enough to cause gravitational collapse and
forms black hole, instead of neutron star. If the pulsar will be not found
in SN 1987A, it may be worth considering this effect seriously.

\placetable{light}

\subsection{Comparison with the solar system abundances \label{comparison}}

\indent

Next, we calculate the total amount of heavy elements in the range
$A = 16-73$ and compare them with solar system abundances.
We note that the informations about initial mass function (IMF),
chemical composition of ejecta for each mass range of the progenitor,
and the ratio of
Type I and type II supernova are necessary when the reproduction
of solar system abundances is attained. In this paper, we can only
make a suggestion for
the degree of impact of axisymmetric explosion on the reproduction of
solar system abundances, because we have calculated only with 
6 $M_{\odot}$ model.
The calculations of explosive nucleosynthesis for wide range of
progenitor's mass are now underway.

We will comment on this analysis. At first, all unstable
nuclei produced in the calculation are assumed to decay to the
corresponding stable nuclei when compared with the solar values.
Secondary, two forms of the mass cut, which are mentioned above, are
used in this analysis to see its influence on the result.

\begin{figure}
\epsscale{1.0}
\plotone{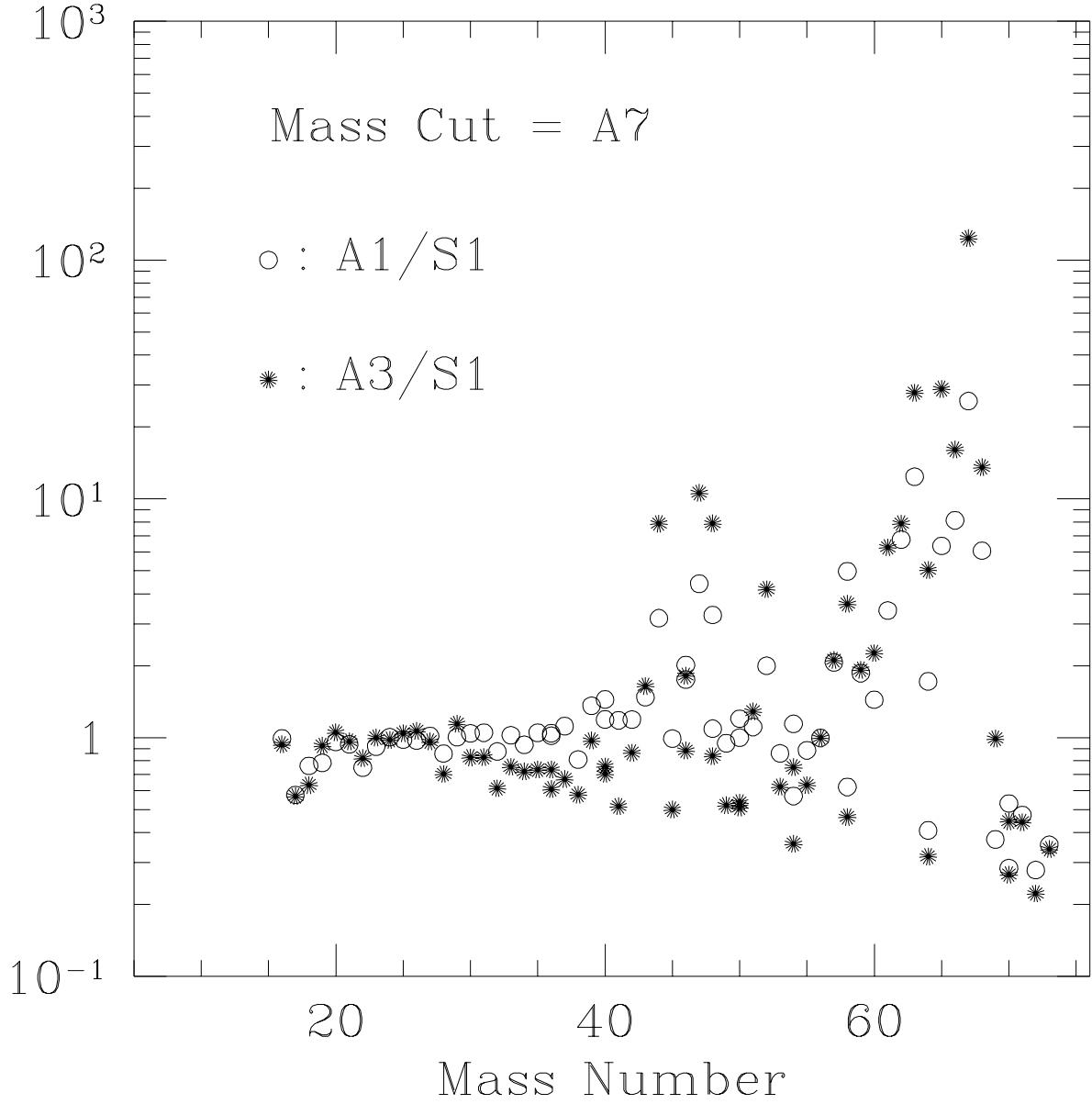}
\figcaption{Comparison of the composition for the mass number range $A=16-73$
normalized by S1 model. The mass cut is set to be A7. Open circles denote the
A1/S1 and dots denote A3/S1. The enhanced nuclei near $A=45$ are
$\rm ^{44}Ca$, $\rm ^{47}Ti$, $\rm ^{48}Ti$, and $\rm ^{52}Cr$, respectively.
\label{fig6}}
\end{figure}

\begin{figure}
\epsscale{1.0}
\plotone{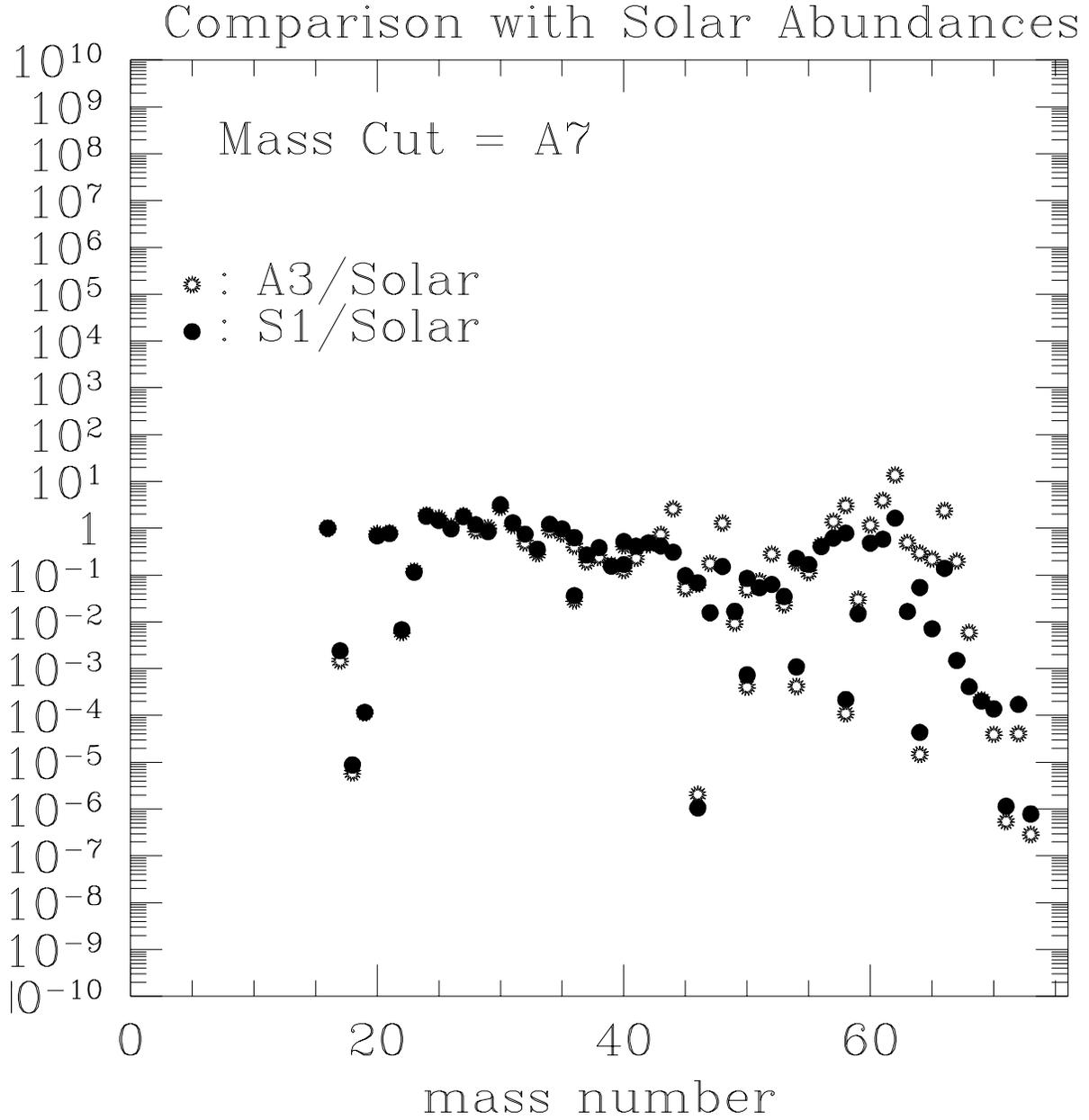}
\figcaption{Comparison of the abundance of each nucleus with the
solar value (normalised to $\rm ^{16}O$).
Open circles and solid points correspond to S1/solar and
A3/solar, respectively. 
\label{fig7}}
\end{figure}

\begin{figure}
\epsscale{1.0}
\plotone{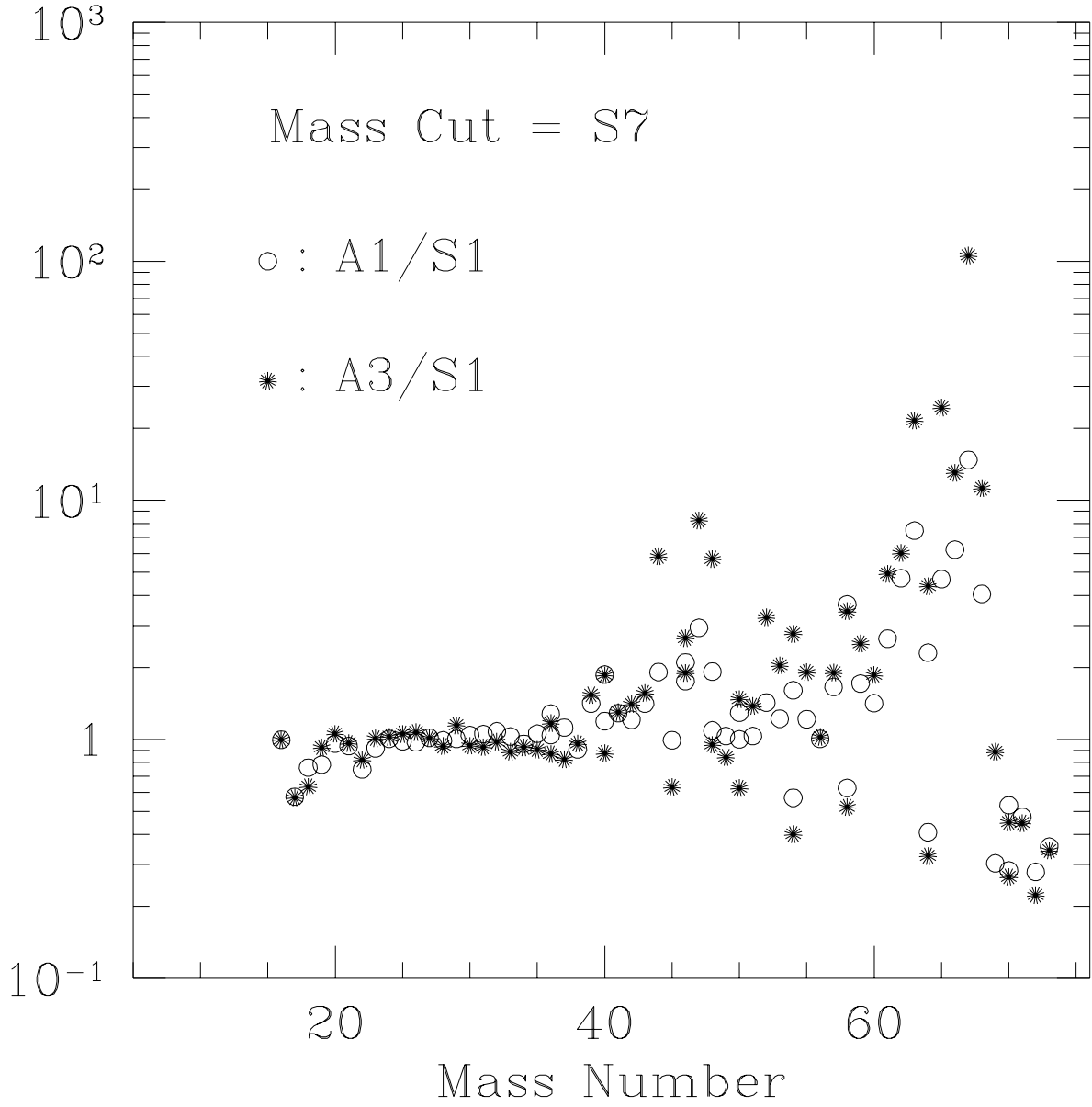}
\figcaption{Same as Fig.6 but for the mass cut of S7.
\label{fig8}}
\end{figure}

Figures ~\ref{fig6} and ~\ref{fig7} show the results
for three models, that is, the S1, A1, and A3 cases for the mass cut
of A7.
Figure ~\ref{fig6} shows the comparison of the composition for $A=16-73$
normalized by the S1 case. Open circles denote the 
A1/S1 and dots denote the A3/S1. Figure~\ref{fig7} 
illustrates the comparison of the abundances of
ejected nuclei with the solar values (normalized at $\rm ^{16} O$).
It is evident from the Figure \ref{fig6}
that the amount of nuclei in the range $A=16-40$ is almost same
among three models and there are two peaks around $A=45$ and $A=65$.
We will comment on these three outstanding feature.
At first, it is the important result that the amount of nuclei is
hardly changes
in the range $A=16-40$, since spherical calculations 
can reproduce well the solar system abundances in this
range (\cite{hashimoto95}). 
Secondary, the enhanced nuclei near $A=45$ are $\rm ^{44}Ca$, $\rm
^{47}Ti$, $\rm ^{48}Ti$, and $\rm ^{52}Cr$, which
are synthesized by the alpha-rich freezeout like $\rm ^{44}Ti$ in
subsection \ref{with1987a}.
We can say that this enhancement is an additional evidence for the more
active alpha-rich freezeout under the axisymmetric explosion. 
Thirdly, the peak around $A=65$ is thought to be made by the strong
shock in the polar region in the axisymmetric explosion, which can
cause nuclear reaction against the coulomb repulsion. Figure
~\ref{fig8} is same as Figure ~\ref{fig6} but for the mass cut of S7.
The peak around A=65 still exists and this suggests the feature of
overproduction of heavy elements around A=65 is fatal one for the
axisymmetric explosion.

We will pay attention to each nuclei individually.
As mentioned in section \ref{intro}, it is reported that $\rm ^{35}Cl,
\; ^{39}K, \; ^{44}Ca$ are underproduced and $\rm ^{58}Ni$ is
overproduced in spherical calculations to date.
The amounts of $\rm ^{35}Cl \; and \; ^{39}K$ produced by the asymmetric
explosion are almost the same as those by the spherical explosion.  
However, $\rm ^{44}Ca$ is produced more and may save the less
production of it in spherical calculation.
For $\rm ^{58}Ni$, it is very sensitive to the $Y_e$ of the
progenitor and position of the mass cut. However, we can say that
neutronization should be also suppressed in the axisymmetric explosion
as seen in section \ref{1987A} to be consistent with the observation.

\placefigure{fig6}
\placefigure{fig7}
\placefigure{fig8}

\section{ Summary and Discussion} \label{summary}

\indent

We have calculated the explosive nucleosynthesis in the supernova of
$6 M_{\odot}$ helium core on the assumption that the explosion is
axisymmetric. We inspected the effect of axisymmetric explosion by
comparing the results with the observation of SN 1987A and solar
system abundances.

As for SN 1987A, it is shown that $\rm ^{44}Ti$ is produced in the
axisymmetric explosion so much as to explain the tail of the light curve.
To put it differently, the degree of the axisymmetric explosion must
be that in our models at least to explain that observation. Although our
forms of the initial shock wave are only assumption, our results
suggest the lower constraint on the degree of the axisymmetric
explosion by the amount of $\rm ^{44}Ti$.
We note that some our results strains the limits set on $\frac {\rm
^{44}Ti}{\rm ^{56}Ni}$ under the spherical explosion
(\cite{woosley91}). We think
this is because the localized energy in the polar region introduce
high entropy and produces much $\rm ^{44}Ti$ there. 
It is difficult to estimate precisely its amount
by the bolometric light curve because of the presence of the bright
optical surroundings and the freeze out effect (\cite{fransson93}).
However, this amount
would be identified from observations by X--ray and $\gamma$--ray
satellites in future. Therefore, the observation of it may
reveal the effects of an axisymmetric explosion.

Moreover, it is reported that the estimated mass of $\rm ^{44}Ti$ by
the observation of
Cas A shows (1.4$\pm$0.4)--(3.2$\pm$0.8)$\times
10^{-4}M_{\odot}$(\cite{iyudin94}). 
Altough its abundance is still controversial,
this value is more than that which is
predicted by spherical calculation (\cite{timmes96}).
Since the form of the Cas A remnant is far from spherical, it may support
our results.

There will be nothing to be complained of if the amount of other
nuclei such as $\rm
^{57}Ni$ and $\rm ^{58}Ni$ are consistent with observations. However,
in the present study, the amount of them tend to be overproduced in
the axisymmetric explosion. It is noted that the $6 M_{\odot}$ helium
core is in itself too neutron-rich in Si layer and the value of $Y_e$ needs to
be modified to explain the observed $\rm ^{58}Ni$ even in the spherical
explosion (\cite{hashimoto95}). We also showed in this paper that if the
value of $Y_e$ in Si layer is modified the amount of $\rm
^{57}Ni$ and $\rm ^{58}Ni$ can be in the range of the observation with 
the amount of $\rm ^{44} Ti$ hardly changed.
It will be necessary to perform our calculation with various
presupernova model to see the dependence of our results on the progenitor's
model, in particular, $Y_e$ distribution.

We also compared the results with solar system abundances. There are
three outstanding features in the axisymmetric explosion. 
One is that the amount of nuclei in the range $A=16-40$ is hardly changed
by the form of the initial shock wave.
The others are that there are two peaks around $A=45$ and $A=65$.
The insensitivity of these nuclei is good for axisymmetric explosions
because of the fact that the spherical calculations thus far can
explain the solar system abundances well in this range (for the spherical case,
see \cite {hashimoto95}). However, we
cannot also solve the problem of the underproduction of $\rm ^{35}Cl$ and
$\rm^{39}K$.
We note $\rm ^{44}Ca$, which is mainly synthesised through the decay
of $\rm ^{44}Ti$, is produced more in the axisymmetric explosion.
This means the axisymmetry of explosion has a positive effect on
explaining the solar values of this nucleus.
On the other hand, the effect of the peak around $A=65$ may be relatively small
since Type I supernovae are chiefly responsible for this mass number range
(e.g., \cite{hashimoto95}). This means, fortunately, the problem of
overproduction of heavy elements in this range may not occur. We are
now calculating explosive nucleosynthesis for wide range of
progenitor's mass and will report on this influence in the near
future. 
We will give an additional comment.
The more production of $\rm ^{44}Ca$ means the more active alpha-rich
freezeout, that is, more helium is remained because of high entropy
per baryon.
We should note that $\rm ^4He$ produced in this region may 
have an effect on the production of $^7 \rm Li$ through the neutrino process.
In addition, the production of $^7 \rm Li$ becomes more efficient 
if axisymmetric radiation of neutrinos (\cite{shimizu94}) causes
the explosion axisymmetric.

We will consider the reliability of our calculations.
As for the mesh resolution, 10 angular zones may seem coarse. However, in the 
calculation of explosive nucleosynthesis, convection will not play an
important role and we think high mesh resolution is not needed.
We are doing more precise calculation using a supercomputer now. We
will estimate the dependence of results on mesh resolution in
future.
In addition, we should explore the sensitivity of results to the initial
condition, such as total explosion energy, the ratio of initial
thermal energy to kinetic energy, and initial radius of the shock wave. 
This is because with either method for the initial shock condition,
that is, the method of depositing energy and of the piston,
the peak temperatures are incorrect in the early history of the shock
(\cite{aufderheide91}).
In the present circumstances, there is no self-consistent way to
produce initial shock wave and the only thing we can do would be to
calculate with various initial conditions. 
Finally, we comment on the dimension of the simulations.
All the simulations presented here are performed in two dimensions since
three dimensional computations require much more computer time and memory.
Two dimensional calculation may limit some flow modes which are
allowed in three dimensional simulations.
As a result of it, the entropy per baryon may be kept higher in the
polar region after the passage of the shock wave, which generates much 
$\rm ^{44}Ti$ in two dimensional calculations. However, we think its
influence will be small since equatorial symmetry will be kept
approximately in the jet-like explosion and 
convection will not play an important role in this study. 
Anyway, we should examine this two dimensional effect in future.
We will estimate the influence of these effects mentioned above and
confirm the credibility of each calculation.

We have done the calculation of the explosive nucleosynthesis under the 
axisymmetric explosion for the first time. 
The feature of axisymmetric explosion may have a lot of attractions 
in addition to explosive nucleosynthesis.
For example, it is shown that a part of $\rm ^{56} \rm Ni$ can be
mixed into the outer layer by Rayleigh-Taylor instabilities in the
jet-like explosion with smaller perturbations compared with spherical
explosion (\cite{yamada91}). 
In addition, the axisymmetric mass cut and jet-like
explosion may be advantageous for ejecting r-process matter. 
The amount of r-process material which is ejected should be very little,
and simultaneously, must be ejected. We think the jet-like explosion
will be one of the mechanism which can explain such characteristic. 
Moreover, some effects such as a rotation of the progenitor 
may show the axisymmetric explosion and it may be essential feature
for presupernova to explode (e.g., \cite{yamada94}, \cite{monchmeyer91}).

This work is the first step for the explosive nucleosynthesis under
the axisymmetric explosion. However, the axisymmetric explosion have
a possibility to solve many mysteries of supernova explosion.
We will calculate axisymmetric explosive nucleosynthesis systematically
for various progenitor models with different masses to
reveal the influence of axisymmetric explosions more clearly.

\acknowledgements
S. N. is grateful to T. Shimizu and Y. Inagaki for a
useful discussion. We are also grateful to E. Skillman for his
kind review of the manuscript. 
M. H. would like to thank M. Arnould and M. Rayet for their
hospitality during his stay at Universit\'e Libre de Bruxelles.
We thank S. Woosley for providing us with the details of his
progenitor models and for his kind comments on this manuscript.
This work has been supported in part by the Grant-in-Aids for
Scientific Research from the Ministry of Education, Science, and Culture
of Japan (07CE2002, 07640386, and 07304033).

\vskip1.0cm

\parskip2pt
\bigskip


\clearpage

\clearpage

\begin{deluxetable}{ccc}
\footnotesize
\tablecaption{Radius of the interface for each layer \label{radius}}
\tablewidth{0pt}
\tablehead{
\colhead{interface} & \colhead{radius [cm]}   & \colhead{radius [$M_{\odot}$]}
} 
\startdata
Fe/Si & $1.5 \times 10^8 $ & $1.4$ \nl
Si/O  & $3.0 \times 10^8 $ & $1.7$ \nl
O/He  & $6.3 \times 10^9 $ & $3.8$ \nl 
\enddata 
\end{deluxetable}

\begin{table*}
\begin{center}
\begin{tabular}{crrrrrrrrrrr}
\tableline
\tableline
Model       & S1  &  A1  &   A2 &  A3  \\
$\alpha$    &  0  & 1/3  &  3/5 &  7/9 \\
$V_p$:$V_e$ & 1:1 & 2:1  &  4:1 &  8:1 \\
\tableline
\end{tabular}
\end{center}

\tablenum{2}
\caption{
Models for the initial shock wave. The first column
shows names for each model. Second is the value of $\alpha$ for each model.
Third is the ratio of the velocity in
the polar region ($\theta = 0 ^{\circ}$) to that in the equatorial
region ($\theta = 90 ^{\circ}$).  \label{model}}

\end{table*}

\begin{table*}
\begin{center}
\begin{tabular}{crrrrrrrrrrr}
\multicolumn{1}{c}{$Y_e$\tablenotemark{a}} & Model &
Mass Cut & Mass of NS &
\multicolumn{1}{c}{$<^{44}\rm Ti / ^{56}\rm Ni>$\tablenotemark{b}}  &
\multicolumn{1}{c}{$<^{57}\rm Ni / ^{56}\rm Ni>$\tablenotemark{c}}  &
\multicolumn{1}{c}{$<^{58}\rm Ni / ^{56}\rm Ni>$\tablenotemark{d}}  \\
\tableline
Case A & S1 & S7 & 1.59$M_{\odot}$ & 0.74 & 1.5 & 2.0 \\
Case A & A1 & S7 & 1.57$M_{\odot}$ & 1.4  & 2.5 & 7.1 \\
Case A & A2 & S7 & 1.56$M_{\odot}$ & 2.2  & 2.8 & 7.5 \\
Case A & A3 & S7 & 1.55$M_{\odot}$ & 4.3  & 2.8 & 6.6 \\
Case A & S1 & A7 & 1.59$M_{\odot}$ & 0.74 & 1.5 & 2.0 \\
Case A & A1 & A7 & 1.57$M_{\odot}$ & 2.4  & 3.2 & 9.8 \\
Case A & A2 & A7 & 1.61$M_{\odot}$ & 3.7  & 3.3 & 8.8 \\
Case A & A3 & A7 & 1.68$M_{\odot}$ & 6.0  & 3.2 & 7.1 \\
Case B & S1 & S7 & 1.59$M_{\odot}$ & 0.76 & 1.5 & 1.5 \\
Case B & A1 & S7 & 1.59$M_{\odot}$ & 1.4  & 1.7 & 1.9 \\
Case B & A2 & S7 & 1.58$M_{\odot}$ & 1.9  & 1.7 & 1.8 \\
Case B & A3 & S7 & 1.57$M_{\odot}$ & 4.0  & 1.8 & 1.6 \\
Case B & S1 & A7 & 1.59$M_{\odot}$ & 0.76 & 1.5 & 1.5 \\
Case B & A1 & A7 & 1.63$M_{\odot}$ & 2.2  & 1.8 & 1.5 \\
Case B & A2 & A7 & 1.75$M_{\odot}$ & 3.5  & 1.8 & 1.3 \\
Case B & A3 & A7 & 1.80$M_{\odot}$ & 6.0  & 1.8 & 0.97 \\
\end{tabular}
\end{center}

\tablenotetext{a}{The way $Y_e$ is modified. See Table caption.}
\tablenotetext{b}{$<^{44}\rm Ti / ^{56}\rm Ni> \equiv [\it X(\rm
^{44}Ti)/ \it X(\rm ^{56}Ni)]/[\it X(\rm ^{44}Ca)/ \it X(\rm^{56}Fe)]_{\odot}$}
\tablenotetext{c}{$<^{57}\rm Ni/ ^{56}\rm Ni> \equiv [\it X(\rm
^{57}Ni)/ \it X(\rm^{56}Ni)]/[\it X(\rm ^{57}Fe)/ \it X(\rm ^{56}Fe)]_{\odot}$}
\tablenotetext{c}{$<^{58}\rm Ni/ ^{56}\rm Ni> \equiv [\it X(\rm
^{58}Ni)/ \it X(\rm^{56}Ni)]/[\it X(\rm ^{58}Ni)/ \it X(\rm ^{56}Fe)]_{\odot}$}

\tablenum{3}
\caption{
$<^{44}\rm Ti/^{56}\rm Ni>$, $<^{57}\rm Ni/^{56}\rm Ni>$, and
$<^{58}\rm Ni/^{56}\rm Ni>$. 
The value of $Y_e$ for $M > 1.607 M_{\odot}$ (=0.494) is artificially
changed to that of $M > 1.637 M_{\odot}$ (=0.499) for Case A. Case B
is same as Case A but for $ M > 1.5 M_{\odot}$. Mass Cut means the
form of it and Mass of NS means baryon mass of the central compact object
(Neutron Star). \label{light}}
\end{table*}

\end{document}